\documentclass[12pt]{iopart}
\usepackage{color}
\usepackage{graphicx}

\begin{document}
\title{Poissonian noise assisted transport in periodic systems}
\author{J Spiechowicz and J {\L}uczka}
\address{Institute of Physics, University of Silesia, Katowice, Poland}
\ead{jerzy.luczka@us.edu.pl}
\begin{abstract}
We revisit the problem of transport of a harmonically driven inertial particle moving in a {\it symmetric} periodic potential, subjected to  {\it unbiased} non-equilibrium generalized white Poissonian noise and coupled to thermal bath. Statistical asymmetry of Poissonian noise is sufficient to induce transport and under presence of external harmonic driving this system exhibits a phenomenon of multiple velocity  reversals. Consequently, one can manipulate the direction of transport just by adjusting the parameters of externally applied forces. 
\end{abstract}
\submitto{\PS}
\maketitle
\section{Introduction}

Transport of Brownian particles in periodic substrates can be controlled by deterministic, external biased force \cite{machura2007}. Less trivial is situation when control is performed by unbiased,  zero-mean force \cite{luczka1999,hanggi2009}. If additionally the unbiased force is random the method of transport control by such random perturbations is non-trivial. In the paper we revisit this problem and study an  inertial particle moving in a  symmetric, spatially periodic  potential. The particle is driven by a simple harmonic force and subjected to both thermal equilibrium  fluctuations and asymmetric Poissonian shot noise \cite{spiechowicz2013}. All forces acting on the particles are of zero mean and directed motion of the particle is induced by asymmetry of the Poissonian noise. We demonstrate that velocity reversal can be detected in the system and show how conveniently manipulate the direction of the Brownian particle velocity. 

The paper is organized as follows. In section 2 we recall details of the model under study. Section 3 is devoted to analysis of the transport characteristics of the Brownian particle. Last but not least, section 3 provides summary and some conclusions.

\section{Model}
In what follows we concentrate on transport of a classical Brownian particle of mass $M$ moving in one-dimensional geometry. Our model  consists of the following elements:  (i) the particle moves  in a \textit{symmetric} spatially periodic potential $V(x) = V(x + L) = \Delta V \sin{(2\pi x/L)}$ of period $L$, (ii) it is driven by an unbiased {\it symmetric} time-periodic force $A\cos{\Omega t}$ with amplitude $A$ and angular frequency $\Omega$, and (iii) is coupled to thermostat of temperature $T$. 
All three elements (i)-(iii) are symmetrical. Therefore the averaged velocity $\langle v \rangle$  of the particle  is zero in the stationary regime. In order to obtain directed transport with $\langle v \rangle \ne 0$, symmetry has to be broken. We introduce the last element which breaks the symmetry, namely (iv) the zero-mean Poissonian shot noise $\eta(t)$. 
Now, the dynamics of the Brownian particle is determined by the Langevin equation in the form \cite{spiechowicz2013}
\begin{equation}
	\label{eq1}
	M\ddot{x} + \Gamma\dot{x} = -V'(x) + A\cos(\Omega t) + \eta(t) + \sqrt{2\Gamma k_BT}\xi(t),
\end{equation}
where  a dot and prime denote differentiation with respect to time $t$ and the Brownian particle's coordinate $x$, respectively. {\it Symmetrical} thermal fluctuations due to the coupling of the particle with the thermostat are modelled by unbiased $\delta$-correlated Gaussian white noise $\xi(t)$ of zero mean and unit intensity, so $\langle \xi(t) \rangle = 0$ and $\langle \xi(t)\xi(s) \rangle = \delta(t-s)$. The parameter $\Gamma$ is the friction coefficient and  $k_B$ is the Boltzmann constant. The noise intensity factor $2\Gamma k_B T$ follows from the fluctuation-dissipation theorem \cite{kubo66,zwan,chaos} and ensures the canonical Gibbs state when $A=0$ and $\eta(t)=0$. All four forces in the rhs of equation (\ref{eq1}) are of zero mean: the mean conservative force $\langle F(x) \rangle = -\langle V'(x) \rangle = 0$ over the space period $L$ and the mean external driving $\langle f(t) \rangle = \langle  A\cos{(\Omega t)} \rangle = 0$ over the time period $T=2\pi/\Omega$. 

The zero-mean non-equilibrium force $\eta(t)$ is chosen in the form of a random sequence of Dirac  $\delta$-shaped pulses with random amplitudes $z_i$ defined in terms of generalized white \textit{Poissonian shot noise}
\begin{equation}
	\label{eq2}
	\eta(t) = \sum_{i=1}^{n(t)}z_i \delta(t - t_i) - \lambda \langle z_i \rangle,
\end{equation}
The random times $t_i$ form a Poisson sequence, i.e., the probability that a sequence of $k$ impulses occurs in the interval $(0,t)$ is given by the Poisson distribution
\begin{equation}
    \label{eq3}
    \mbox{Pr}\{n(t) = k\} = \frac{(\lambda t)^k}{k!}e^{-\lambda t}.
\end{equation}
The parameter $\lambda$ is the mean number of $\delta$-pulses per unit time. 
The amplitudes $\{z_i\}$ of the $\delta$-pulses are independent random variables distributed according to a common probability density $\rho(z)$. As an example, we consider the density $\rho(z)$ in the exponential form 
\begin{equation}
	\label{eq4}
    \rho(z) = \zeta^{-1} \theta(z) \exp{(-z/\zeta)},
\end{equation}
where $\theta(z)$ denotes the Heaviside step function. Hence, the amplitudes take only positive values, $z_i > 0$. Their moments, according to (\ref{eq4}), are given by the relation
\begin{equation}
    \label{eq5}
    \langle z_i^k \rangle = k! \zeta^k, \quad k = 1, 2, 3..
\end{equation}
In particular, the mean value $\langle z \rangle = \zeta$ and it gives interpretation of the parameter $\zeta$. The process $\eta(t)$ is white noise of zero mean and the Dirac delta autocorrelation function, namely,  
\begin{equation}
    \label{eq6}
    \langle \eta(t) \rangle = 0, \quad \langle \eta(t)\eta(s) \rangle = 2\lambda\zeta^2\delta(t-s).
\end{equation}
The last equation defines the Poissonian noise intensity 
\begin{equation}
    \label{eq7}
    D_P = \lambda\zeta^2.
\end{equation}
Moreover, we assume that thermal equilibrium noise $\xi(t)$ is uncorrelated with non-equilibrium noise $\eta(t)$, so
\begin{equation}
    \label{eq9}
    \langle \xi(t)\eta(s) \rangle = \langle \xi(t) \rangle \langle \eta(s) \rangle = 0.
\end{equation}
Such white Poissonian noise commonly occurs in various micro-structures \cite{czernik1997} and is characterized by a \textit{temporal asymmetry}, i.e. sharp $\delta$-pulses of zero duration are followed by a constant negative bias which lasts over an exponentially distributed waiting time. In \cite{luczka1995}, it has been demonstrated that the white noise nature of such \textit{additive}, \textit{temporally asymmetric} fluctuations is sufficient to induce directed transport in periodic structures, in the presence and in the absence of an internal asymmetry. This should be contrasted with the case of \textit{temporally symmetric} white Poissonian noise (i.e. with equally probable positive and negative amplitudes $\rho(z) = \rho(-z)$) which is able to generate a net macroscopic velocity only if the reflection symmetry of the periodic structure is broken, hence only for the so called ratchet systems \cite{hanggi2009}. A directed transport emerges non-trivially if backward as well as forward transitions drive the particle and if no balancing between them takes place. The white Poissonian noise characterizes non-equilibrium fluctuations and therefore detailed balance does not hold. However, this alone does not guarantee a non-zero averaged particle velocity. The necessary condition is a source of \textit{statistical asymmetry}. In the case of white Poissonian noise this asymmetry has its roots in the non-vanishing odd higher order cumulants, namely \cite{hanggi1996}
\begin{eqnarray}
    \label{eq10}
    C_{2n+1}(t_1, ..., t_{2n+1}) &= \langle \eta(t_1)...\eta(t_{2n+1}) \rangle \nonumber\\ &= \lambda(2n+1)!\zeta^{2n+1}\delta(t_1 - t_2)...\delta(t_{2n} - t_{2n+1}),
\end{eqnarray}
where $n = 1, 2, ...$ The reader should always remember that the statistics of the process $\eta(t)$ is given by all cumulants and according to the above equation it is clearly not symmetric. As a result, the backward and forward transitions are not equal and the directed transport can emerge.

Let us now introduce the dimensionless form of (\ref{eq1}). This can be done in several different ways. Here we propose to use the period $L$ as a length scale and for time the scale $\tau = L\sqrt{M/\Delta V}$ \cite{spiechowicz2013}. Then  (\ref{eq1}) can be rewritten in the form
\begin{equation}
    \label{eq12}
    \ddot{\hat{x}} + \gamma\dot{\hat{x}} = - \hat{V}'(\hat{x}) + a\cos(\omega \hat{t}) + \hat{\eta}(\hat{t}) + \sqrt{2\gamma D_G}\, \hat{\xi}(\hat{t}),
\end{equation}
where $\hat{x} = x/L$ and $\hat{t} = t/\tau$. Other re-scaled dimensionless parameters are the friction coefficient $\gamma = \tau \Gamma/M$, the amplitude $a = LA/\Delta V$ and  the angular frequency $\omega = \tau \Omega$ of the time-periodic driving. The rescaled  potential $\hat{V}(\hat{x}) = V(L\hat{x})/\Delta V = \sin(2\pi \hat{x})$ possesses the unit period:  $\hat{V}(\hat{x}) = \hat{V}(\hat{x} +1)$. The rescaled zero-mean thermal noise has intensity $D_G = k_BT/\Delta V$  and the auto-correlation function $\langle \hat{\xi}(\hat{t})\hat{\xi}(\hat{s}) \rangle = \delta(\hat{t} - \hat{s})$. Similarly, the re-scaled zero-mean Poissonian white shot noise is  $\delta$-correlated: $\langle \hat{\eta}(\hat{t})\hat{\eta}(\hat{s}) \rangle = 2\hat{D}_P\delta(\hat{t} - \hat{s})$  with intensity $\hat{D}_p = \hat{\lambda} \langle \hat{z}_i^2 \rangle /2$, where $\hat{\lambda} = \tau \lambda$ and $\hat{z}_i = z_i/\sqrt{M\Delta V}$. Hereafter, we will use only  dimensionless variables and shall omit the notation "hat" in all quantities appearing in equation (\ref{eq12}).

The most prominent transport quantity for the system (\ref{eq12})  is the average dimensionless velocity $\langle \dot x(t)\rangle$ of the Brownian particle. In the long time limit, it can be presented in the form of a series of all possible harmonics, namely,  
\begin{equation}
	\label{asym}
	\lim_{t\to\infty} \langle {\dot x(t)} \rangle =  \langle v \rangle + v_{\omega}(t) + v_{2\omega}(t) + \dots, 
\end{equation}
where $\langle v \rangle $ is a dc (time-independent) component and $v_{n \omega}(t)$ are time-periodic functions which time average over a basic period $\omega$ are  zero. In this case the dc component $\langle v \rangle$ is obtained after averaging over the temporal period of the driving and the corresponding ensemble-average \cite{jung1993}, namely, 
\begin{equation}
	\label{eq11}
	\langle v \rangle = \lim_{t\to\infty} \frac{\omega}{2\pi} \int_{t}^{t+2\pi/\omega} \prec \dot{x}(s) \succ \; ds,
\end{equation}
where $\prec \cdot \succ$ denotes the average over noise realizations. In the deterministic case ($D_G = D_P = 0$), an additional averaging over initial conditions must be performed.
\section{Transport properties}
In order to obtain the relevant transport characteristics we have performed comprehensive numerical simulations of driven Langevin dynamics determined by equation (\ref{eq12}). Details of the employed numerical scheme can be found in \cite{kim2007,grigoriu2009}. We have set the time step to be $0.0005 \cdot 2\pi/\omega$ and for the  initial conditions $\{x(0),\dot{x}(0)\}$ we used a uniform distribution over the interval $[0, 1]$ and $[-2,2]$, respectively. Quantity of interest was ensemble averaged over $10^3 - 10^4$ different trajectories which evolved over $10^3 - 10^4$ periods of the external AC driving.  All numerical calculations were done by use of a CUDA environment implemented on a modern desktop GPU. This scheme allowed for a speed-up of a factor of the order $10^3$ times as compared to a common present-day CPU method \cite{januszewski2009}.

Let us start our analysis of transport properties of inertial Brownian particles described by (\ref{eq12}) with short comment on the impact of the  Poissonian noise parameters $\lambda$ and $D_P$ on its stochastic realizations. The reader can find detailed discussion on this topic in \cite{spiechowicz2013}. Here, we only mention two limiting cases. The first extreme regime is when both $\lambda$ and $D_P$ are large. Then the Brownian particle is very frequently kicked by large $\delta$-pulses. Since the distance between two successive Poissonian arrival times is very short there are only a few moments when merely the negative valued bias of the process acts on the system. On the contrary, when both $\lambda$ and $D_P$ are small, then the particle is very rarely kicked by weak $\delta$-pulses. It also means that there are long periods of time in which the system is exposed only to the action of the negative bias of the non-equilibrium noise.
\begin{figure}[h]
    \centering
    \includegraphics[width=0.49\linewidth]{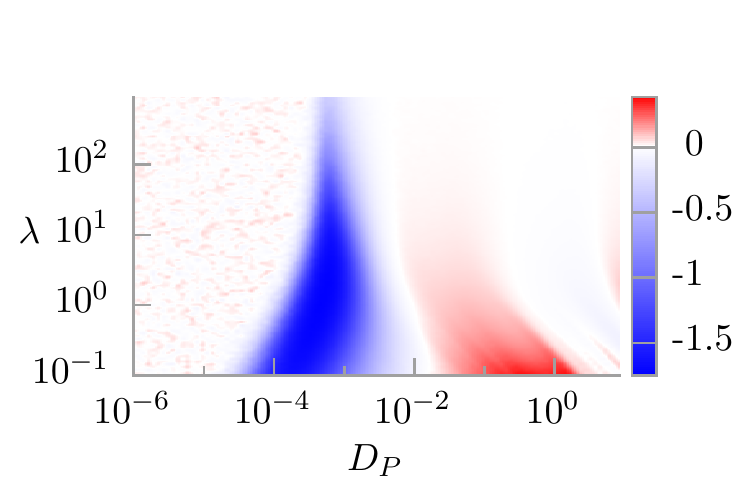}
    \caption{Regions of  positive (red) and negative (blue) averaged velocity $\langle v \rangle$ in the parameter plane $\{D_P, \lambda\}$ of the white Poissonian noise intensity $D_P$ and the spiking rate $\lambda$. Other parameters are: $a = 10.8$, $\omega = 3.77$, $\gamma = 1.04$ and $D_G = 0$.}
    \label{fig1}
\end{figure}

\begin{figure}[h]
    \centering
    \includegraphics[width=0.49\linewidth]{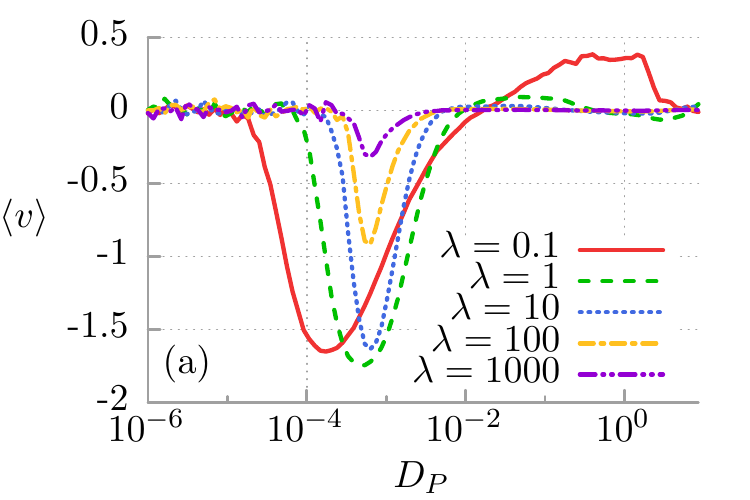}
    \includegraphics[width=0.49\linewidth]{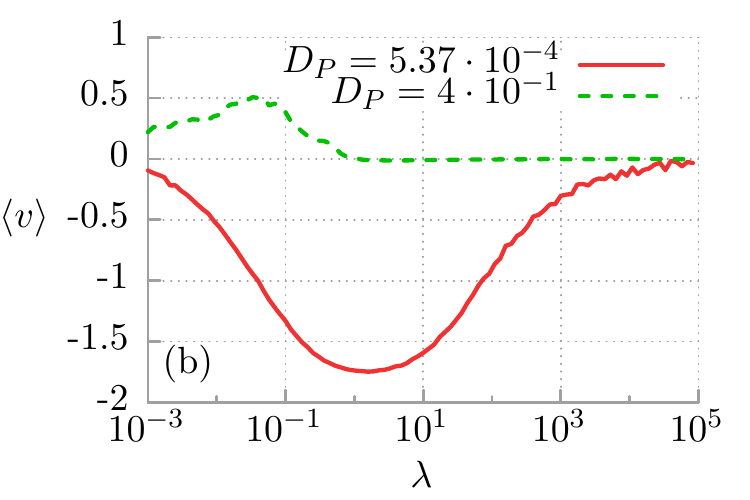}
    \caption{Averaged velocity of the Brownian particles $\langle v \rangle$ is presented as a function of the white Poissonian noise intensity $D_P$ and the spiking frequency $\lambda$ in the panel (a) and (b), respectively. Other parameters are the same as in figure \ref{fig1}.}
    \label{fig2}
\end{figure}

The equation given by (\ref{eq12}) has a multidimensional parameter space, namely $\{\gamma, a, \omega, \lambda, D_P, D_G\}$. To eliminate one of them we first look at the dynamics in the case  when $D_G = 0$. We limit our considerations to positive driving amplitudes $a$ noting that the system (\ref{eq12}) is invariant under changes of sign of $a$. It is sufficient to investigate low and moderate driving frequencies $\omega$ because under very fast positive and negative oscillations of driving the average velocity $\langle v \rangle$ cannot be induced. This procedure leaves us with 5-dimensional parameter space the detailed exploration of which is still hopeless numerically even for the powerful currently available personal GPU supercomputers. Therefore we present selected transport regimes which exhibit interesting behaviour.
Figure \ref{fig1} depicts the influence of variation of the Poissonian noise intensity $D_P$ and the frequency $\lambda$ of the Dirac $\delta$-pulses on the velocity $\langle v \rangle$. 
Transport is negligibly small for both small and large values of the Poissonian noise intensity $D_P$ regardless of the magnitude of the spiking rate $\lambda$. There are only two clear distinguished islands corresponding to the negative and positive velocity, respectively. Moreover, one can observe that for particular, fixed non-equilibrium noise intensities $D_P$ the direction of transport is constant irrespective of the variation of the spiking frequency of the $\delta$-kicks. Probably the most interesting is the existence of a wide window of the rates $\lambda$ for which one can conveniently manipulate the direction of transport process just by adjusting the noise intensity $D_P$. These findings are confirmed in figure \ref{fig2} where sample cuts of the previous panel are presented. In particular, panel (a) illustrates the dependence of the average velocity $\langle v \rangle$ on the Poissonian noise intensity $D_P$ for several selected values of the spiking rate $\lambda$. According to the previous statement one can observe there the phenomenon of multiple velocity  reversals \cite{jung1996,mateos2000,mateos2003,kostur2000} (see the case $\lambda = 1$). Moreover, it is seen that an increase of the frequency $\lambda$ has destructive impact on the modulus of the directed velocity $\langle v \rangle$. The non-equilibrium noise of very small and very large intensity cannot induce noticeable velocity. In panel (b) the same transport characteristic is depicted but versus the spiking rate $\lambda$ for two  Poissonian noise intensities $D_P$ corresponding to the minimum and maximum of the curve presented in figure \ref{fig2}(a). It is worth to note that for both fixed noise intensities there is an optimal frequency $\lambda$ to generate a non-zero velocity. However, its direction is constant regardless of the value of the spiking rate $\lambda$.

\begin{figure}[h]
    \centering
    \includegraphics[width=0.49\linewidth]{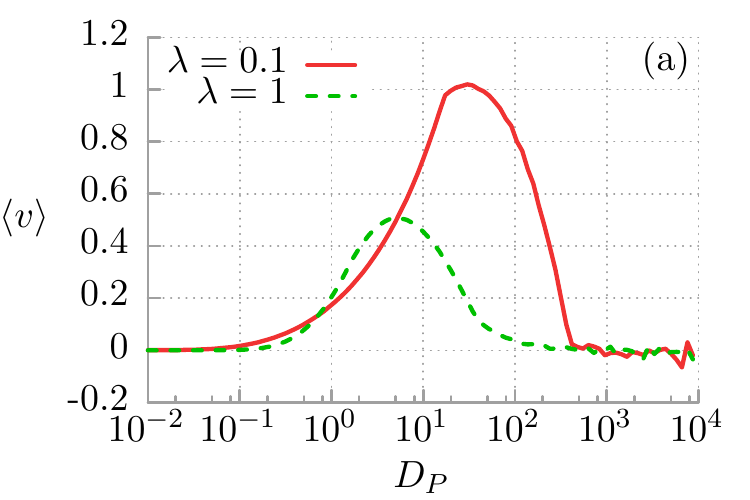} \\
    \includegraphics[width=0.49\linewidth]{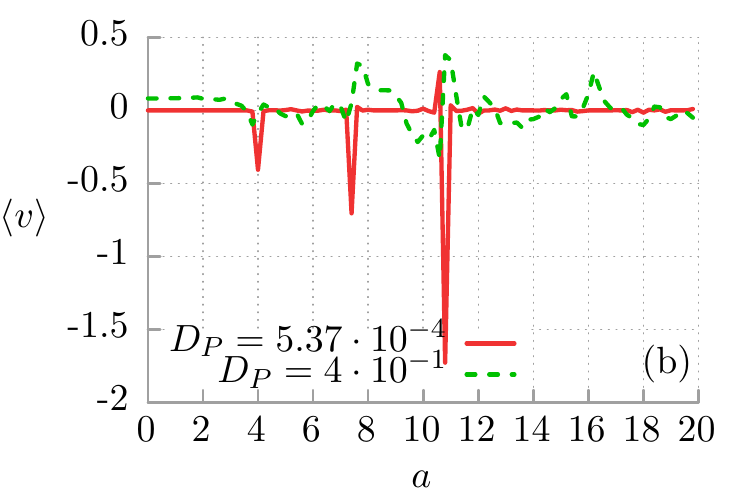}
    \includegraphics[width=0.49\linewidth]{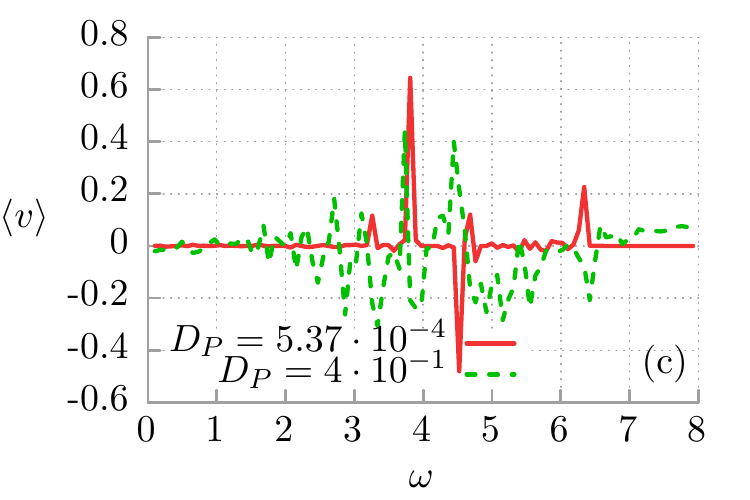} 
    \caption{Role of the harmonic driving  in transport process. Panel (a): the asymptotic long time average velocity $\langle v \rangle$ as a function of the white Poissonian noise intensity $D_P$ in the absence of the harmonic driving force $a=0$. Panel (b): the dependence of the particle velocity on the amplitude $a$ of the driving. Panel (c): the influence of the variation of the frequency $\omega$ of the harmonic force on the average directed velocity $\langle v \rangle$. Other parameters are the same as in figure \ref{fig1}.}
    \label{fig3}
\end{figure}

The next three panels are devoted to the role of the harmonic force in the transport process. In particular, the first one depicts the same characteristic as in figure \ref{fig2}(a) but in the absence the driving, i.e. when  $a=0$. One should note that in the line of earlier remarks for sufficiently large $D_P$ the white Poissonian noise is able to solely induce finite asymptotic long time average velocity $\langle v \rangle$. There is an optimal region of $D_P$ in which the stationary velocity is maximal.  However, for the case $a=0$ there is no phenomenon of the velocity reversal. Consequently, in this regime the harmonic driving force plays a crucial role and allows for steering of the direction of transport. Its significance is further analysed in the rest two plots of figure \ref{fig3}. Panel (b) and (c) illustrates the average velocity $\langle v \rangle$ versus the amplitude $a$ and the frequency $\omega$ of the harmonic driving, respectively. One can see there that the amplitude $a$ can serve as a convenient parameter to manipulate the direction of transport. The same conclusion also holds true in the case of the frequency $\omega$ of the harmonic driving force (c.f. figure \ref{fig3}(c)). Furthermore, the dependence of the average velocity of the Brownian particle on the parameters of the driving $a\cos{(\omega t)}$ often depicts the resonance like behaviour: small variation of their magnitude can lead to rapid changes of the sign and value of the velocity. For small non-equilibrium noise intensities $D_P \to 0$ the velocity $\langle v \rangle$ is negligibly small in both limiting cases of $\omega \to 0$ and $\omega \to \infty$.

\begin{figure}[h]
    \centering
    \includegraphics[width=0.49\linewidth]{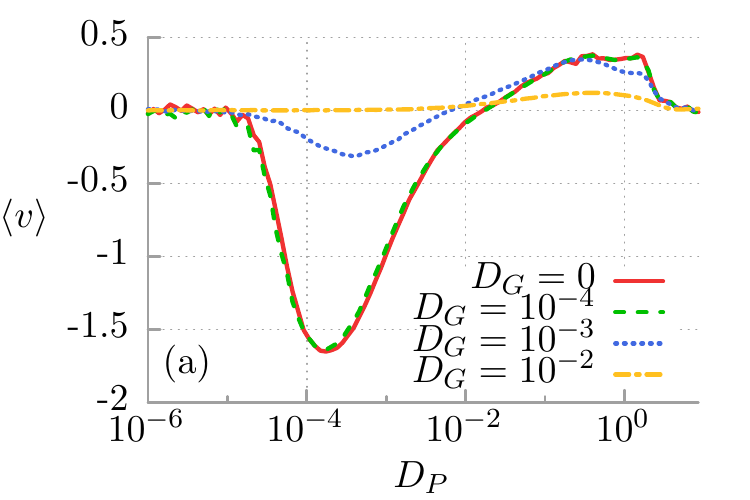} 
    \includegraphics[width=0.49\linewidth]{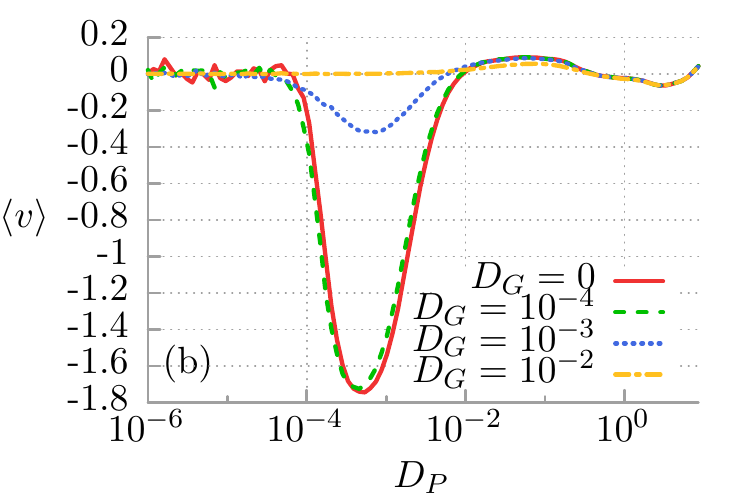} \\
    \includegraphics[width=0.49\linewidth]{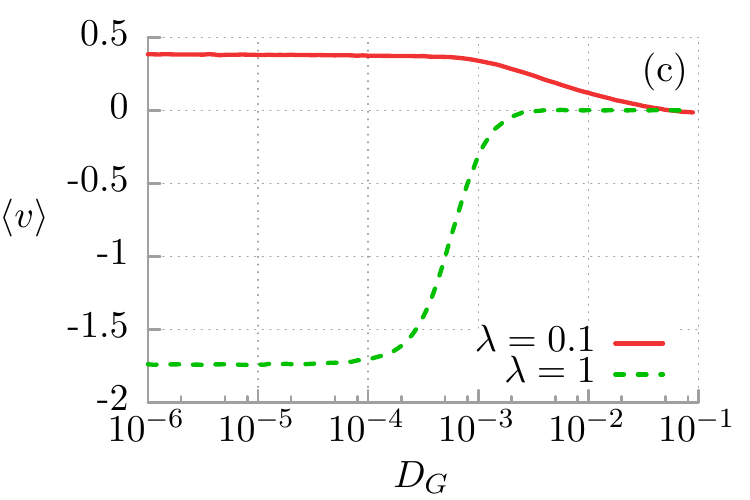} 
    \caption{The role of the coupling of the particle to the heat bath in the observed transport process. Panel (a) and (b): the asymptotic long time average velocity $\langle v \rangle$ as a function of the white Poissonian noise intensity $D_P$ for several different thermal noise intensities $D_G$ and two spiking rates $\lambda = 0.1$ and $\lambda = 1$, respectively. Panel (c): the dependence of the particle velocity on the thermal fluctuations intensity $D_G$ for $\lambda = 0.1$, $D_P = 4 \cdot 10^{-1}$ (red, solid line) and $\lambda = 1$, $D_P = 5.37 \cdot 10^{-4}$ (green, dashed line). Other parameters are the same as in figure \ref{fig1}.}
    \label{fig4}
\end{figure}

We now analyse the impact of coupling of the particle to thermostat on the transport process, see figure \ref{fig4},  where the average velocity $\langle v \rangle$ is plotted for different values of thermal noise intensity $D_G \propto T$. Temperature $T$ has a smoothing effect on the plots, erasing the finer details of the structures visible in previous figures. This is to be expected, because the introduction of thermal noise causes additional random transitions between coexisting basins of attraction. The common opinion on thermal noise says that it has negative impact on the transport processes. This case is realized here as well. A careful inspection of figure \ref{fig4}(c) reveals that indeed an increase of thermal noise intensity leads to a decrease of the observed particle velocity. It is an example of destructive influence of thermal noise on the transport process. However, both presented regimes are quite temperature resistant as the average velocity $\langle v \rangle$ starts to drop significantly only for high intensities of thermal fluctuations. The non-zero velocity is caused by stochastic, complex chaotic dynamics and  even at zero temperature $D_G = 0$, the average velocity is non-zero. 
\section{Summary}
With this study we  analysed transport properties of  inertial Brownian particles which move in a \emph{symmetric} periodic potential and are subjected to both a \emph{symmetric}, unbiased time-periodic external force and a \emph{temporally asymmetric} generalized white Poissonian noise. First, we demonstrated that the white noise nature of such temporally asymmetric fluctuations is sufficient to generate the directed transport of the under-damped system  in the spatially periodic structure. Second, under presence of the external harmonic driving it is possible to observe the phenomenon of multiple velocity reversals. One can conveniently manipulate the direction of the particle velocity  by tuning of the parameters of unbiased time-periodic external force and the white Poissonian noise intensity. We have also elucidated that this transport phenomenon emerges as a result of statistical asymmetry of the non-equilibrium noise and is quite robust with respect to the variation of temperature. 

Finally, let us remind that the Langevin equation {(\ref{eq1}) has similar form as an equation of motion for the phase difference $\Psi=\Psi(t)$ between the macroscopic wave functions of the Cooper pairs on both sides of the Josephson junction. The quasi-classical dynamics of the resistively and capacitively shunted Josephson junction is well known in the literature as the Stewart-McCumber model \cite{stewart,mccumber,junction,kautz}. Therefore our results can readily be experimentally tested with an accessible setup consisting of a single Josephson junction device operating in its quasi-classical regime.

\ack
J S is supported by the FORSZT project co-financed by EU from the European Social Fund and  J {\L} is supported by the NCN grant DEC-2013/09/B/ST3/01659.

\end{document}